\UseRawInputEncoding
\documentclass[pr, twocolumn, preprintnumbers, showpacs, footnoteadded, superscriptaddress,longbibliography]{revtex4-1}
\pdfoutput=1

\usepackage{amssymb}
\usepackage{graphicx}
\usepackage{amsmath}
\usepackage{hyperref}
\usepackage{tensor}
\usepackage{epstopdf}
\usepackage{extarrows}
\usepackage{graphicx,subfigure}

\newcommand{\td}{\text{d}}
\newcommand{\te}{\text{e}}

\newcommand{\be}{\begin{equation}}
\newcommand{\ee}{\end{equation}}
\newcommand{\bea}{\begin{eqnarray}}
\newcommand{\eea}{\end{eqnarray}}
\newcommand{\ba}{\begin{eqnarray}}
\newcommand{\ea}{\end{eqnarray}}

\def\o{\omega}

\begin{document}
\title{\boldmath Superradiantly tability analysis on dyonic stringly black hole}
\author{Jia-Hui Huang}
\email{huangjh@m.scnu.edu.cn}
\affiliation{Guangdong Provincial Key Laboratory of Quantum Engineering and Quantum Materials,
School of Physics and Telecommunication Engineering,
South China Normal University, Guangzhou 510006, China}
\affiliation{Guangdong-Hong Kong Joint Laboratory of Quantum Matter,
Southern Nuclear Science Computing Center, South China Normal University, Guangzhou 510006, China}
\affiliation{Guangdong Provincial Key Laboratory of Nuclear Science, Institute of quantum matter,
South China Normal University, Guangzhou 510006, China}
\author{Mu-Zi Zhang}
\author{Tian-Tian Cao}
\author{Yi-Feng Zou}
\affiliation{Guangdong-Hong Kong Joint Laboratory of Quantum Matter,
Southern Nuclear Science Computing Center, South China Normal University, Guangzhou 510006, China}
\affiliation{Guangdong Provincial Key Laboratory of Nuclear Science, Institute of quantum matter,
South China Normal University, Guangzhou 510006, China}
\author{Zhan-Feng Mai}
\email{zhanfeng.mai@gmail.com}
\affiliation{Center for Joint Quantum Studies and Department of Physics, School of Science, Tianjin University, Yaguan Road 135, Jinnan District, 300350 Tianjin, P. R. China}

\begin{abstract}
The black hole superradiance phenomena states that when a scalar wave perturbation is properly scattering off a charged or rotating black hole, the energy of the reflected scalar wave may be amplified. If this amplification process can occur back and forth through certain confining mechanism, it will lead to strong instability of the black hole, which is so-called "black hole bomb". In this paper, the superradiant stability is investigated for dyonic black holes in string theory. Although the electric charged black hole in string theory has been proved superrdiantly stable, it is found that the dyonic black hole is more unstable than a RN black hole due to the magnetic charge. Furthermore, by our analysis on the effective potential outside the event horizon, we give the parameters region associated with the superradiance stability on dyonic stringly black hole.
\end{abstract}

\maketitle
\flushbottom

\noindent

\section{Introduction}
The (in)stability problem of black holes is an interesting topic in black hole physics. Regge and Wheeler\cite{wheeler1957} proved that the spherically symmetric Schwarzschild black hole is stable under perturbations. The stability problems of rotating or charged black holes are more complicated due to the significant effect of superradiance. Superradiance effect can occur in both classical and quantum scattering processes
\cite{Manogue1988,Greiner1985,Cardoso2004,Brito:2015oca,Brito:2014wla}. When a charged bosonic wave is impinging upon a charged rotating black hole, the wave reflected by the event horizon is amplified if the wave frequency $\omega$ satisfied the following superradiant regime
\begin{equation}\label{suprad re}
\omega < m\Omega  + e\Phi,
\end{equation}
where $e$ is the charge of the bosonic wave, $m$ is the azimuthal number, $\Omega$ is the angular velocity of black hole horizon and $\Phi$ is the electromagnetic potential of the black hole \cite{P1969,Ch1970,M1972,Ya1971,Bardeen1972,Bekenstein1973,Ya1972}. This means that when the incoming wave is scattered, the wave extracts rotational energy from rotating black holes and electromagnetic energy from charged black holes. The Eq.~\eqref{suprad re} then is called superradiance condition. According to the black hole bomb mechanism proposed by Press and Teukolsky\cite{PTbomb}, if there is a mirror between the black hole horizon and space
infinity, the amplified wave has been trapped as well as can be reflected back and forth between the mirror and the black hole, growing  exponentially, resulting in the superradiant instability of the black hole.

The superradiant properties of different kinds of black holes have been extensively studied in the literature, such as Hawking radiation (for a review, see \cite{Brito:2015oca}). It has been illustrated that the angular momentum of black hole will trigger the superradiant instability. A rotating black hole can be superradiantly unstable caused by a massive scalar field when the parameters of the black hole and the scalar filed are in certain  parameter spaces. In this case, the mass of the scalar field acts as a natural mirror. A lot of work has been done to identify these parameter spaces, e.g. \cite{Strafuss:2004qc,Konoplya:2006br,Cardoso:2011xi,Dolan:2012yt,Hod:2012zza,Hod:2014pza,Aliev:2014aba,Hod:2016iri,Degollado:2018ypf,Huang:2019xbu}.
A rapidly spinning black hole that is impinged upon by a complex and massive vector field is discussed in \cite{East:2017ovw,East:2017mrj}. The superradiant instability of rotating black holes in curved space is also reported
\cite{Cardoso:2004hs,Cardoso:2013pza,Zhang:2014kna,Delice:2015zga,Aliev:2015wla,Wang:2015fgp,Ferreira:2017tnc,Konoplya:2013rxa,Kokkotas:2015uma}.

It is well known that superradiant instability can also occur for charged black holes. When there is a mirror or a cavity outside a Reissner-Nordstrom (RN) black hole horizon, this black hole is superradiantly unstable in certain parameter spaces \cite{Herdeiro:2013pia,Li:2014gfg,Degollado:2013bha,Sanchis-Gual:2015lje,Fierro:2017fky,Gonzalez:2017gwa}. However, in flat background, RN black holes have been proved to be superradiantly stable against charged massive scalar perturbation\cite{Hod:2013eea,Huang:2015jza,Hod:2015hza,DiMenza:2014vpa}. It is pointed that when the parameters of a RN black hole and a charged massive scalar field satisfy superradiant conditions, there is no effective trapping potential outside the black hole horizon, which acts as a mirror to reflect the superradiant modes \cite{Huang:2015jza}. When charged black holes are in curved backgrounds, such as (anti-)de Sitter space, these backgrounds provide  natural reflecting boundary conditions and support the existence of superradiant instability \cite{Wang:2014eha,Bosch:2016vcp,Huang:2016zoz,Gonzalez:2017shu,Zhu:2014sya}. The discussion of black hole superradiant property has also been extended to the analogue of RN black hole in string theory  and it has been shown that stringy RN black hole is  superradiantly stable against charged massive scalar perturbation\cite{Li:2013jna,Konoplya:2011qq}. But when  a mirror is introduced, superradiant modes exist and the stringy RN black hole becomes unstable \cite{Li:2014xxa,Li:2014fna,Li:2015mqa}.

In addition, magnetic field can trigger superradiant instability for a black hole. In \cite{Konoplya:2007yy,Konoplya:2008hj}, the authors use an approximation method to show that when a scalar field is propagating on the Ernst background, the magnetic field will induce an effective mass $\mu_{eff}\propto B$($B$ is the magnetic field strength) for the scalar field, leading to the superradiant instability. In a further work \cite{Brito:2014nja}, the authors considered Kerr-Newman black holes immersed in a uniform magnetic field and showed that  the magnetic field can confine scalar perturbations leading to long-lived modes, which trigger superradiant instabilities.

Based on our illustration above, it is thus worth asking a question whether a series of asymptotic flat dyonic (with both electric charge and magnetic charge) stringly black hole is superradiantly stable against a massive charged perturbative scalar field. In our previous work \cite{Huang:2015jza}, it was demonstrated that the effective potential outside the outer horizon have not potential well when both the superradiance condition and bound state condition were imposed, we thus showed that electrically charged RN black holes are superradiantly stable under massive charged scalar perturbation. However, in this paper, we will consider the superradiant behavior of the dyonic black holes under a charged massive scalar perturbation. For our case, the metric of the dyonic black hole is similar to a electrically charged RN black hole,but pthe magnetic field also leads to the superradiant instability of the dyonic black holes and scalar perturbation system.

\section{The Superradiance Condition of Dyonic Black Hole}
As our demonstration in \cite{Huang:2015jza}, the electric charged of RN black hole in asymptotic flat will not trigger the superradiance instability under the superradiance condition $\omega < \omega_{c} = \frac{q Q_e}{r_+}$ in which $Qe$ and $r_+$ denote the electric charge and event horizon of the RN black hole respectively. A worth-answering question is whether the magnetic charge associated with black hole will change the superradiance condition. From the view of classical electromagnetic dynamic, the Lorentz force associated to the static magnetic field does not do work as well as increase(decrease) the kinetic energy carried by a given charged particle. Therefore, we suppose a conjecture that the magnetic charged carried by the dyonic black hole will not chage the superradiance condition. In following, we shall prove this conjecture.

At first, we give a brief introduction on the scalar perturbation related to the stringly dyonic black hole. The action reduced from low energy effective theory of string theory in four dimension, written in the Einstein-frame, is given (here we adopt the natural unit as $G=c=\hbar=k=1$)
\begin{equation}\label{action2}
S=\frac{1}{16 \pi }\int_{M} \td^4 x \sqrt{-g} \left(R-\frac{1}{2}\left(\partial \phi \right)^2 - \te^{-2 \phi}F^{\mu\nu}F_{\mu\nu} \right),
\end{equation}
where $F_{\mu\nu}=\partial_\mu A_\nu - \partial_\nu A_{\mu}$ and $\phi$ denote the strength of Maxwell field and dilaton field respectively. By performing variation on the action, we will obtain the equation of motion
\begin{eqnarray}
&& R_{\mu\nu}-\frac{1}{2}g_{\mu\nu}R=\frac{1}{2}\nabla_\mu \phi \nabla_\nu \phi \cr
&&+ 2 \te^{-2 \phi}F^\mu_{\rho}F_{\nu}{}^\rho-\frac{1}{2}g_{\mu\nu}\left(\frac{1}{2}(\partial \phi)^2 - \te^{-2\phi}F^{\mu\nu}F_{\mu\nu}\right)\, , \\
~ \cr
&& \nabla_\nu \left(\te^{-2 \phi}F^{\mu\nu}\right)=0 \, , \\
~ \cr
&& \square \phi + 2 \te^{-2 \phi}F^{\mu\nu}F_{\mu\nu} \phi=0.
\end{eqnarray}
The theory Eq.~\eqref{action2} admits a series of stringly dyonic black hole solutions with the ADM mass $M$, electric charge $Q_e$ and magnetic charge $Q_m$\cite{Horowitz:1992jp,Sen:1992fr}. Specifically, the metric of this class of stringly dyonic black hole can be written as follow,
\begin{eqnarray}
&& \td{s^2} =  - \frac{\Delta}{\rho(r)}\td{t^2}+ \frac{\rho(r)}{\Delta}\td{r^2}+\rho(r)(\td\theta^2 +\sin^2\theta \td\varphi^2) \, , \\
~ \cr
&& A_\mu = -\frac{Q_e}{r} \left(\td t \right)_\mu - Q_m \cos \theta \left(\td \varphi \right)_\mu \, , \\
~ \cr
&& \te^{2 \phi}=\frac{r-r_0}{r+r_0} \, ,
\end{eqnarray}
where
\begin{eqnarray}
&& \Delta=(r-r_+)(r-r_-) \, , \rho(r)=r^2-r_0^2\,  \\
&&r_0=\frac{Q^2_m-Q^2_e}{2M} \, , \\
&&r_{\pm}=M\pm \sqrt{M^2+r^2_0-Q^2_m-Q^2_e} \, .
\end{eqnarray}
%


To proceed, we then derive the master equation associated with the perturbative charged scalar field $\Psi$ under the dyonic black hole background. Due to the static and spherical dyonic black hole background, the equation of motion associated with $\Psi$ can reduce to a radial equation in 1 dimension in frequency domain. The dynamics of the charged massive scalar field is described by the Klein-Gordon (KG) equation
\begin{equation}\label{scaeom}
\left(({\nabla ^\nu } - iq{A^\nu })({\nabla _\nu } - iq{A_\nu }) - {\mu
^2}\right)\Psi  = 0,
\end{equation}
where $q$ and $\mu$ are the charge and rest mass of the charged scalar field. The nonzero vector field components are $A_t=-Q_e/r$ and $A_\varphi=-Q_m \cos\theta$, describing the electric and magnetic charges related to the black hole\cite{Sen:1992fr}.

Applying to the variables separation method in spherical black hole background, we take the anstanz of $\Psi$ as
\begin{equation}\label{scaan}
\Psi(t,r,\theta,\varphi)=R(r)Y(\theta)\te^{im\varphi}\te^{-i\omega t}\, ,
\end{equation}
where $m$ is azimuthal harmonic index and $\omega$ is the frequency of the scalar perturbation. Based on the periodic condition related to $\varphi$,
\begin{equation}\label{period}
\varphi \to \varphi + 2 \pi \, ,
\end{equation}
the invariance of exponential complex function with respect to $\varphi$, $\te^{i m \varphi}=\te^{im \varphi} \te^{i 2 m \pi }$, requires the discrete spectrum of the azimuthal harmonic index, namely $m=0, \pm 1 \, \pm 2 \cdots$. We then focus on investigating the angular function $Y(\theta)$,  Plugging the anstanz Eq.~\eqref{scaan} into the equation of motion Eq.~\eqref{scaeom}, the angular equation with respect to $Y(\theta)$ gives
\begin{eqnarray}\label{angfun}
&&\frac{1}{\sin\theta}\partial_\theta(\sin\theta\partial_\theta Y) \cr
&&+(\lambda_{l m}-\frac{m^2+q^2 Q_m^2+2m q Q_m\cos\theta}{\sin^2\theta})Y=0 \, ,
\end{eqnarray}
where $\lambda_{l m }$ is so-called separation constant, as well as usually set to $\lambda=l(l+1)$. By Performing variable transformation as $\eta=\cos\theta$ for convenience, we arrive that the angular equation Eq.~\eqref{angfun} will deform to a class of  the Fuchs equation
\begin{eqnarray}\label{Fuchs equation}
&&\frac{\td^2Y}{\td\eta^2}+(\frac{1}{\eta-1}+\frac{1}{\eta+1})\frac{\td Y}{\td \eta}+\left(-\frac{(m+q Q_m)^2}{2(\eta-1)} \right. \cr
~\cr
&&+ \left.\frac{(m-q Q_m)^2}{2(\eta+1)}-\lambda \right)\frac{Y}{(\eta-1)(\eta+1)}=0,
\end{eqnarray}
In general, the physical boundary condition requires that the angular function $Y(\theta)$ should be regular at $\theta = 0, \pi$, corresponding to $\eta = \pm 1$, we then choose one branch of solution associated with Fuchs equation \eqref{Fuchs equation},
\begin{eqnarray}\label{angfun2}
Y(\eta)=&&C(1-\eta)^{\frac{1}{2}(m+q Q_m)}(1+\eta)^{\frac{1}{2}(m-q Q_m)} \cr
~\cr
        &&{}_2F_{1}\left(\alpha, \beta; \gamma,  \frac{1-\eta}{2}\right) \, ,
\end{eqnarray}
where $C$ is a normalization constant. The parameters $\alpha,\beta$ and $\gamma$  are given by
\begin{equation}
\left\{ \begin{array}{l}
{\alpha=m+ \frac{1}{2}(1+\sqrt{1+4\lambda})}\\
~\cr
 {\beta=m+ \frac{1}{2}(1-\sqrt{1+4\lambda})}\\
 ~\cr
 {\gamma=1+m-qQ_m}
 \end{array} \right.
\end{equation}
The regular condition of $Y(\eta)$ in the region $-1\leqslant\eta\leqslant1$ also requires that
\begin{itemize}
\item{ $\beta$ is a negative integer, corresponding to the hypergeometric function reducing to hypergeometric polynomial which must be regular in $\eta = \pm 1$, giving that the separation constant $\lambda=l(l+1)$ and $l> |m|$.}
\item{In addition to regularity of requirement associated with the hypergeometric polynomial, the regularity of the factor $(1-\eta)^{\frac{1}{2}(m+q Q_m)}(1+\eta)^{\frac{1}{2}(m-q Q_m)}$ in \eqref{angfun2} implies the following two inequalities,
\begin{equation}\label{mqQ}
m>q Q_m\, , \quad l(l+1)>q^2 Q_m^2 \, .
\end{equation}
}
\end{itemize}

It is remarkable to note that different from the cases of RN black hole and the electrically charged black hole in string theory, the solution of angular equation $Y(\eta)$ will be no longer hypergeometric function due to the interaction between the charged and magnetic field of the black hole. Through the regular analysis on the hypergeometric function, the regularity associated with the angular function $Y(\theta)$ gives the lower bounds of the angular quantum numbers, namely $l(l+1) > q^2 Q_m^2$. It means that due to the interaction between the charged scalar and the magnetic field, $l$ will not vanish, implying that the perturbative particle must spin around the black hole under the magnetic field semi-classically.

Finally, we arrive the radial equation of the perturbative charged scalar field in frequency domain,
\begin{equation}\label{ridial eq}
\Delta \frac{\td}{\td r}\left(\Delta \frac{\td R(r)}{\td r}\right)+U(r)R(r)=0 \, ,
\end{equation}
where
\begin{equation}
U(r)=\left( \omega-\frac{q Q_e}{r} \right)^2 \rho(r)^2-\Delta \left(l\left(l+1 \right)-q^2 Q_m^2 - \mu^2 \rho(r) \right)
\end{equation}

In order to study whether the superradiance condition of the dyonic black hole against the massive charged perturbative scalar field will be changed by the static magnetic charge, one need to impose both the ingoing and outgoing wave boundary condition at the infinity, as well as the ingoing wave boundary condition near the horizon region. That indicates that a beam of charged scalar field propagate from spatial infinity to the black hole horizon and then is scattered by the black hole and a part of this beam is reflected to the infinity while the other part was absorbed by the black hole. To achieve that, it is convenient to define the tortoise coordinate $y$ as
\begin{equation}
\frac{\td y}{\td r}=\frac{r^2}{\Delta},
\end{equation}
and a new radial function as ${\tilde R}=r R$, the
radial wave equation \eqref{ridial eq} can be written in following standard wave function form
\begin{equation}\label{radto}
\frac{\td^2\tilde R}{\td y^2}+\tilde U\tilde R=0 ,
\end{equation}
where the potential
\begin{equation}\label{ubar}
\tilde
U=\frac{U}{r^4}+\frac{2\Delta^2}{r^6}-\frac{\Delta}{r^5}\frac{d\Delta}{dr}.
\end{equation}
To observe the behavior of the radial function $\tilde{R}$, one can asymptotically perform Taylor expansion at both near horizon and the spatial infinity, specifically have
\begin{equation}
\lim_{r \to {r_+}} \tilde U = \left(\omega-\frac{qQ_e}{r_+} \right )^2 \left(r_+^2-r_0^2 \right)^2, ~~ \lim_{r \to +\infty } \tilde U = \omega^2 - {\mu ^2} \, .
\end{equation}
The radial function $\tilde{R}$ then can be solved as
\begin{equation}\label{asyso}
 \tilde R \varpropto \left\{
\begin{aligned}
&{\cal T} \te^{i  k_+ y }\, ,  \quad  &y \to -\infty \, ,  \cr
&{\cal L} \te^{i \sqrt{\omega^2-\mu^2}y } + {\cal R} {\te^{-i\sqrt{\omega^2-\mu^2}y}}   \, , \quad  &y\to +\infty \, ,
\end{aligned} \right.
\end{equation}
where the wave number in the near horizon region $y \to -\infty$ is
\begin{equation}
k_+ = \pm \frac{1}{r_+}\left(\omega - \frac{q Q_e}{r_+} \right)\left(1-\left(\frac{r_0}{r_+} \right)^2 \right) \, ,
\end{equation}
and also the complex number ${\cal L}$, ${\cal R}$ and ${\cal T}$ are denoted as the incident amplitude, reflection amplitude and the transmission amplitude respectively. Remarkably, the sign of $\omega$ and $k_+$ is decided by the requirement associated with the propagating direction denoted by the group velocity. In general, we require the wave propagate from the spatial infinity $y \to \infty$ to the black hole horizon $y \to -\infty$, implying that $\frac{\partial \omega}{\partial k_+} <0$. We thus choose the negative sign. On the other hand, the ${\cal R}$ and ${\cal T}$ are decided by the explicit shape of the potential $\tilde{U}$. To obtain the superradiance condition, we evaluate the conserved Wronskain associated with the solution $\tilde{R}$ and its complex conjugate $\tilde{R}^{*}$ at both boundary, corresponding to the conserved current density, then have
\begin{equation}
|{\cal R}|^2=\left|{\cal L}\right|^2- \frac{\left(\omega - q Q_e/r_+\right)\left(r_+^2-r_0^2 p\right)}{\omega^2 - \mu^2}\left|{\cal T}\right|^2 \, .
\end{equation}
One can easily observe that when
\begin{equation}\label{supcon}
0 < \omega < \frac{q Q_e}{r_+}\,
\end{equation}
$|{\cal R}|^2 > |{\cal L}|^2$, namely the reflection amplitude is larger than the incident one, indicating that the superradiance phenomena occurs. On the other hand, the above illustration also give the superradiance condition Eq.~\eqref{supcon}. One can easily see that the superradiance condition of dyonic black hole is the same as the case of stringly RN black hole\cite{Li:2013jna}, even the RN black hole in Einstein theory\cite{DiMenza:2014vpa}. Therefore, we have proved our conjecture that the static magnetic field in dyonic black hole in string theory will not effect the superradiance condition. It is remarkable to note that the superradiance condition Eq.~\eqref{supcon} which is important for us to investigate the superradiance instability of the stringly dyonic black hole.

\section{Superradiant Stability analysis of dyonic black hole}
In the last section, we proved that the magnetic field will not effect the superradiance condition $\omega < q Q_e /r_+$ in which $q$, $Q_e$ and $r_+$ are denotes the charge of the perturbative scalar field, the electric charge and the location of the event horizon of the dyonic black hole respectively. In other words, if one inject a beam of perturbative charged scalar field whose the frequency satisfying the superradiance condition $0 < \omega < q Q_e/r_+$, only the electric energy of the dyonic black hole, rather than the magnetic energy, will be extracted by the charged scalar field due to the superrdiance phenomena. The next question we need to answer is whether the dyonic black hole in string theory is superradiantly stable once the superradiance mode of the perturbative charged scalar field has been triggered on. In Ref.~\cite{Li:2013jna,DiMenza:2014vpa}, it has been proved that the pure electric charged black hole either in the Einstein gravity theory or string theory is superradiantly stable against a perturbative charged scalar field. Therefore, it is very interesting to investigate whether the magnetic charge will trigger the superradiant instability in the stringly dyonic black hole. In addition to the superradiance condition, another necessary condition to trigger the superradiance instability of a given black hole is that there exist a trapping well associated with the effective potential outside the event horizon to trap the superradiance mode. If there does exist a trapping well of the effective potential, a bound state of the perturbative scalar field will arise and the superradiance amplification of such a bound state will lead to instability of the black hole, which is so-called the "black hole bomb" mechanic.

To achieve to investigate whether there exist a trapping well triggering the superradiant instability of the stringly dyonic black hole, we at first consider the case of $Q_e = Q_m$. In this case, the derivative function of effective potential, of which the numbers of roots indicates the existence of the trapping well, owe a more simpler polynomial form to analyze. Therefore, we adopt analytical method to study the associated superradiant instability. However, when taking $Q_e \not = Q_m$ into account, one will find that the number of roots of the derivative function related to the effective potential cannot be given efficiently by analytical analysis. Therefore, we turn to investigate the existence of the trapping well of the effective potential outside the horizon by numerical method.

\subsection{The Bound State Condition}
From our illustration above, we study the superradiant stability on dyonic black hole through analysing the existence of trapping well of the potential in term of the radial equation outside the horizon to trap the bound state of $\Psi$. Therefore, we firstly need to give a bound state condition. Practically, To obtain the appropriate effective potential as well as deform the radial equation Eq.~\eqref{ridial eq} to a Schr\"{o}dinger-like equation, it is convenient to define a new radial function $\psi$,
\begin{equation} \label{psiradial}
\psi=\Delta^{\frac{1}{2}}R \, .
\end{equation}
The radial equation Eq.~\eqref{ridial eq} can be written as the following Schr\"{o}dinger-like equation
\begin{equation}
\frac{\td^2\psi}{\td r^2}+(\omega^2-V)\psi=0 \, ,
\end{equation}
where the effective potential is
\begin{equation}
V=\omega^2+\frac{1}{\Delta^2} \left(\Delta-U-(r-M)^2\right) \, .
\end{equation}

Remarkably, in general the radial equation as well as its effective potential is investigated in tortoise coordinate. However, the effective potential given by the radial equation in tortoise coordinate, namely $\tilde{U}$ in Eq.~\eqref{ubar} is not appropriate for our investigation on superradiant instability on dyonic black hole since the effective potential in tortoise coordinate $\tilde{U}$ is effectively approach to constants, rather than zero, in $y \to \pm \infty$ region. The first radial function $\tilde{R}$ then is free propagator in the near horizon region $y \to \infty$. Nevertheless, from the definition of the second radial function $\psi$, namely Eq.~\eqref{psiradial}, one can find that $\psi$ is effectively reduce to zero at the event horizon since $\Delta|_{r=r_+}=0$, corresponding to the bound state condition. To match the bound state condition at the spatial infinity region, we need to analyze the asymptotic behavior of the potential $V$. We have
\begin{equation}\label{rprmV}
V \to -\infty \, , \quad  r \to r_{\pm}
\end{equation}
as well as
\begin{equation}\label{Vinfty}
V \to \mu^2+\frac{-4M\omega^2+2qQ\omega+2M\mu^2}{r}+{\cal O} \left(\frac{1}{r^2} \right), \quad  r \to +\infty.
\end{equation}
The leading order of $V$ implies that the radial equation in terms of $\psi$ will reduce to
\begin{equation}
\frac{\td^2\psi}{\td r^2}+(\omega^2-\mu^2)\psi=0 \, , \quad r \to \infty
\end{equation}
Interpreted $\psi$ as a wave function in bound state after superradiance amplification locally, the bound state condition at spatial infinity requires that
\begin{equation}
\omega<\mu \, ,
\end{equation}
corresponding the exponential decay mode associated with $\psi$ as $\psi|_{r \to \infty} \varpropto \te^{-\sqrt{\mu^2-\omega^2} r}$ at the spatial infinity. Therefore, we finally obtain two necessary condition for us to investigate the existence associated with the trapping well to trap the bound state mode for the potential $V$ outside the horizon, namely the superradiance condition as well as the bound state condition,
\begin{equation}
\omega < \frac{q Q_e}{r_+} \, , \quad \omega < \mu .
\end{equation}

\subsection{Analytical Analysis: The case of $Q_e = Q_m$}
To proceed, we turn to study the general shape of the potential $V$, which can generally depend on six parameters $\{M, Q_e, Q_m, \omega, \mu, q, l\}$ in which the first three parameters are given by the dyonic black hole and the last four parameters denote the property of the perturbative charged scalar field $\Psi$. Naively we consider the case of $Q_e=Q_m=Q$. Recall the subleading order of $V$ given in Eq.~\eqref{Vinfty}, it is easy to prove that
\begin{eqnarray}\label{2plusroot}
&-4M\omega^2+2qQ\omega+2M\mu^2  \cr
~\cr
>&-4M\omega^2+2qQ\omega+2M\omega^2  \quad &(\omega < \mu)  \cr
~\cr
=&2M\omega\left(\frac{qQ}{M}-\omega \right) \quad &(r_+ > M)\cr
~\cr
>&2M\omega \left(\frac{qQ}{r_+}-\omega \right)>0 \quad &(\omega< q Q/r_+) \, ,
\end{eqnarray}
The asymptotic behavior of $V$, Eq.~\eqref{Vinfty} and Eq.~\eqref{2plusroot}, implies that the potential $V$ has at least one maximum in the region $r_-<r<r_+$ and  one maximum outside the  event horizon ($r>r_+$).

In practice, if there was only one maximum outside the event horizon for $V$, there will be no trapping well associated with $V$ to trap the bound state of $\Psi$, indicating that the dyonic black hole will be superradiantly stable. In following we will give a class of region related to the parameter space to illustrate the superradiant stability for the dyonic black hole by analysing the number of root associated with the derivative function of $V$. Since it can be concluded that the dyonic black hole is superradiantly stable when the potential $V$ has only one maximum outside the horizon. According to some basic conclusion related to calculus, it indicates that one can find the regions of parameters to illustrate the superradiant stability on dyonic black hole under some constrains that the derivative function of $V$ has only one root outside the horizon.

Through the analysis on the asymptotic behavior of $V$, it has been concluded that there at least exist one maximum of $V$ when $r_- < r_+$ and $r > r_+$ respectively. For convenience, we define a new radial variable $z=r-r_-$. The derivative function with respect to the $V$ then be given
\begin{equation}
\frac{\td V}{\td z} =  \frac{-2}{\Delta^3}\left( \left(\Delta-U-\left(z+r_--M \right)^2 \right)\frac{\td \Delta}{\td z}+\frac{1}{2}\Delta \frac{\td U}{\td z} \right).
\end{equation}
It is also easy to conclude that there exist at least two roots of the numerator of $\frac{\td V}{\td z}$ when $z > 0$ since the translation of $r$ will not change the number of root associated with the numerator of $V'(z)$. Practically, the numerator of $V'(z)$ can be written in a polynomial form in terms of $z$
\begin{equation}\label{Vd}
f(z)=a z^4+b z^3+c z^2 +d z+e \, ,
\end{equation}
where
\begin{eqnarray}
a &=& -2M\mu^2+4M\omega^2-2qQ\omega \, , \\
~ \cr
b &=&4 \left(8 M^2 - 6 M r_+ + r_+^2 \right)\omega^2-4 q Q \left(5 M - 2 r_+ \right)\omega  \cr
  &-& 2 \left(l + l^2 - 2 q^2 Q^2 + \left(6 M^2 - 6 M r_+ + r_+^2 \right)\mu^2 \right) \, , \\
~\cr
c &=&12 \left(2M-r_+ \right)^3\omega^2-18qQ \left(2M-r_+ \right)^2\omega \cr
  &-&6 \left( \left(M-r_+ \right)l \left(l+1 \right) -3Mq^2Q^2\right.\cr
  &+& \left.2r_+q^2Q^2+\mu^2 \left(M - r_+ \right) \left(2 M-r_+ \right)^2 \right)\, , \\
~ \cr
d &=&4 \left(4 M - 3 r_+ \right) \left(2 M - r_+ \right)^3\omega^2 \cr
  &-&4 q Q \left(7 M - 5 r_+ \right) \left(2 M - r_+ \right)^2\omega \cr
  &-& 4 \left(2 M^2 - 3 M r_+ + r_+^2 \right)^2\mu^2 \cr
  &+& 4 \left(7 M^2 - 9 M r_+ + 3 r_+^2 \right)q^2Q^2  \cr
  &-& 4 \left(l \left(l+1 \right)-1 \right) \left(M - r_+ \right)^2 \, , \\
~ \cr
e &=& 2\left( r_{-}-r_{+} \right)r^2_- \left(qQ-\omega r_- \right)^2+\frac{1}{2}\left(r_{-}-r_{+}\right)^3 \, .
\end{eqnarray}
The four roots of $f(z)$ , denoted by $z_1, z_2, z_3, z_4$, should satisfy the following two identities,
 \begin{eqnarray}\label{prod}
&& z_1 \times z_2 \times z_3 \times z_4=\frac{e}{a} \, , \\    \label{sum}
~\cr
&&z_1 \times z_2+z_1 \times z_3+z_1 \times z_4 + \cr
&&z_2 \times z_3+z_2 \times z_4+z_3 \times z_4 =  \frac{c}{a} \, .\label{sum1}
 \end{eqnarray}
Based on some general algebraic quaternary equation properties, in principle, these four roots are in complex plane. According to our discussion on Eq.~\eqref{2plusroot}, There already have been two real roots. It is easy to see that $V'(z)$ will only have two real roots since if there is complex root among $z_1, z_2, z_3, z_4$, its complex conjugate is also a root. Therefore in the following discussion, we naively suppose a worse case that all four roots are real. Recall the superradiant conditions $\o < qQ/r_+$ as well as the bound state condition $\o <\mu$, one can easily prove that
$a<0$ and $e<0$. As we illustrate above, the asymptotic behaviors of the effective potential indicates that there are at least two maxima for the effective potential $V(z)$ when $z>0$, corresponding to two positive roots $z_1, z_2$ of $V'(z)=0$. The equation Eq.~\eqref{prod} implies the another two roots $z_3, z_4$ are both positive or negative. If $z_3, z_4$ are both negative, then the $V'(z)$ will only have two real roots outside the inner horizon, implying that there is no trapping well outside the event horizon and the black hole is thus superradiantly stable. From equation \eqref{sum}, we conclude that \emph{$c/a <0$ (i.e. $c>0$) is a sufficient condition for negative $z_3, z_4$ and therefore a sufficient condition for
superradiantly stable dyonic black holes}.

In following, we are focusing the parameter regions of dyonic black hole and scalar field where $c>0$. Coefficient $c$ can be treated as a quadratic polynomial of $\o$, $c=c(\o)$ and the coefficient of the quadratic term is positive. The intercept of $c(\o)$ is
\begin{eqnarray}
&&c(0)=-6\left((M-r_+)l(l+1)- 3Mq^2Q^2 \right.\cr
~\cr
&&\left.+2r_+q^2Q^2+\mu^2(M - r_+)(2 M-r_+)^2\right).
\end{eqnarray}
The discriminant of the quadratic equation with respect to $\omega$, namely $c(\omega)=0$, is
\begin{eqnarray}
\Delta_c
= &&36 (2 M-r_+)^3 \left(8 l^2 (M-r_+)+8 l (M-r_+)+32 \mu ^2 M^3 \right.\cr
~\cr
&&-64 \mu ^2 M^2 r_+ - 6 M q^2 Q^2+40 \mu ^2 M r_+^2 \cr
~\cr
&&\left. +7 q^2 Q^2 r_+-8 \mu ^2 r_+^3\right).
\end{eqnarray}
In the following, we shall discuss more detail in two cases for $c>0$, finding out a class of relevant parameter regions to give the superradiantly stable dyonic black hole. For convenience, we define the following dimensionless quality
\begin{equation}
x=\frac{r_+}{M}\, .
\end{equation}

\begin{itemize}
\item { \textbf{Case I:  $\Delta_c <0 $}

In the case of $\Delta_c < 0$, it is easy to prove that $c > 0$ is always holding. Recall the fact that $2M > r_+>M$ as well as $\Delta_c <0 $, it is equivalent to
\begin{eqnarray}
&&8 l^2 \left(M-r_+ \right)+8 l \left(M-r_+ \right)+32 \mu ^2 M^3  -64 \mu ^2 M^2 r_+ \cr
~\cr
&&-6 M q^2 Q^2+40 \mu ^2 M r_+^2+7 q^2 Q^2 r_+-8 \mu ^2 r_+^3<0 \, ,
\end{eqnarray}
which can also be rewritten as
\begin{equation}\label{negDeltaC}
\mu^2>\frac{q^2Q^2}{r_+^2}\frac{x^2 ( 7 x-6)}{8 ( x-2)^2 ( x-1)}-\frac{l(l+1)}{r_-^2}\, .
\end{equation}
Recall definition of $r_+$, we have $1<x<2$.  When the parameters of black hole and dyonic black hole satisfy Eq.~\eqref{negDeltaC}, the dyonic black hole is superradiantly stable against the scalar perturbation. Two examples of the effective potential in this case are shown in Fig.~\ref{Riemann1}
\begin{figure}[hbtp]
\centering
\includegraphics[width=0.4\textwidth]{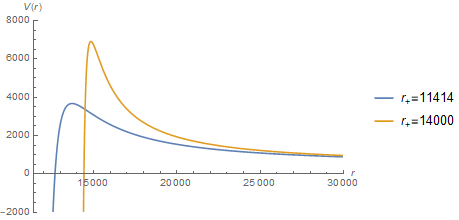}
\caption{Two examples of the effective potential in case I. The black hole mass are chosen as
$r_+=11414$ and $r_+=14000$ for the orange curve and blue curve respectively. The other parameters are chosen as $l=\omega=5, \mu=20, q=15, Q=7000.$  }\label{Riemann1}
\end{figure}

}

\item{\textbf{Case II-I: $\Delta_c >0$, Requiring $0<\omega<\omega_- $ then $c>0$. }

In this case, denoting the two roots of $c(\omega)=0$ as $\omega_\pm$, we firstly consider the case that $0<\omega<\omega_- $, indicating that $c>0$. Explicitly, we have
\begin{equation}
\omega_\pm=\frac{18qQr_-^2\pm\sqrt{\Delta_c}}{24 r_-^3}.
\end{equation}
The condition $\Delta_c >0$ requires that $\mu$ satisfy
\begin{equation}\label{deltaMu}
\mu^2<\frac{q^2Q^2}{r_+^2}\frac{x^2 ( 7 x-6)}{8 ( x-2)^2 ( x-1)}-\frac{l(l+1)}{r_-^2} \, .
\end{equation}
Recall the superradiance condition, namely $\omega < qQ/r_+$, one can find that $0<\omega<\omega_- \Rightarrow c>0$ must satisfy if requiring that $qQ/r_+ <\omega_-$,. Therefore, we follow the illustration above and consider the constraint $qQ/r_+ <\omega_-$, which explicitly is
\begin{eqnarray}
&&qQ/r_+ < \frac{18qQr_-^2-\sqrt{\Delta_c}}{24 r_-^3}=\frac{3qQ}{4r_-}-\frac{\sqrt{\Delta_c}}{24 r_-^3} \cr
~\cr
&&\Leftrightarrow \boxed{\frac{\sqrt{\Delta_c}}{24 r_-^3} < \frac{3qQ}{4r_-}-qQ/r_+} \, .
\end{eqnarray}
Then the above inequality is equivalent to
\begin{eqnarray}
&&\frac{3qQ}{4r_-}-qQ/r_+>0~(r_+/r_->4/3)\, , \\
~\cr
&&18qQr_-^2r_+-24 r_-^3qQ>r_+\sqrt{\Delta_c}\, .
\end{eqnarray}
The second inequality will give that
\begin{eqnarray}
&&r_+^2 (2 M - r_+)^2\mu^2-16 M^2 q^2 Q^2 \cr
~\cr
&&+ 20 M q^2 Q^2 r_+ + (l(l+1) - 7 q^2 Q^2) r_+^2>0,
\end{eqnarray}
which can be simplified as
\begin{eqnarray}\label{poDeltaC}
\mu^2 &>& \frac{q^2Q^2}{r_+^2}\frac{7r_+^2-20Mr_++16M^2}{r_-^2}-\frac{l(l+1)}{r_-^2} \cr
~\cr
&=& \frac{q^2Q^2}{r_+^2}\frac{7x^2-20x+16}{(x-2)^2}-\frac{l(l+1)}{r_-^2} \, .
\end{eqnarray}
One can check that the above condition  is consistent with Eq.~\eqref{deltaMu},
\begin{eqnarray}
&&\frac{q^2Q^2}{r_+^2}\frac{7x^2-20x+16}{(x-2)^2}-\frac{l(l+1)}{r_-^2} \cr
&&<\mu^2<\frac{q^2Q^2}{r_+^2}\frac{x^2 ( 7 x-6)}{8 ( x-2)^2 (x-1)}-\frac{l(l+1)}{r_-^2} \, .
\end{eqnarray}
Therefore, together with the constraints on mass and charge ratio of dyonic black hole, we then obtain a following superradiantly stable parameter region of the scalar and dyonic black hole system in case II-I. We show the plot associated with the effective potential in Fig.~\ref{case2}.
\begin{figure}[hbtp]
\centering
\includegraphics[width=0.4\textwidth]{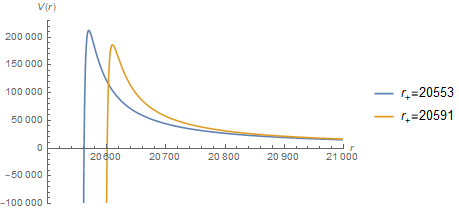}
\caption{Two examples of the effective potential in case II-I. The black hole mass are chosen as
$r_+=20553$ and $r_+=20591$ for the orange curve and blue curve respectively. The other parameters are chosen as $l=\omega=5, \mu=20, q=15, Q=7000.$  }\label{case2}
\end{figure}
\begin{eqnarray}
&&\frac{q^2Q^2}{r_+^2}\frac{7x^2-20x+16}{(x-2)^2}-\frac{l(l+1)}{r_-^2} \cr
~\cr
&&<\mu^2<\frac{q^2Q^2}{r_+^2}\frac{x^2 ( 7 x-6)}{8 ( x-2)^2 (x-1)}-\frac{l(l+1)}{r_-^2} \, .
\end{eqnarray}
with
\begin{equation}
\frac{r_+}{r_-}>\frac{4}{3} \quad \text{or} \quad  \left(\frac{Q}{M}\right) < \frac{2\sqrt{6}}{7}\, .
\end{equation}
}

\item{\textbf{Case II-II: Under $\Delta_c >0$, Requiring $\omega>\omega_+$ then $ c>0$ }

Recall the superradiance condition, $\omega < qQ/r_+$. the requirement of $c > 0$ will give $\omega_+<\omega<qQ/r_+$. Therefore, in following we consider the constraint $qQ/r_+ >\omega_+$. Explicitly, we have
\begin{eqnarray}
&&qQ/r_+ >\frac{18qQr_-^2+\sqrt{\Delta_c}}{24 r_-^3}=\frac{3qQ}{4r_-}+\frac{\sqrt{\Delta_c}}{24 r_-^3} \cr
&&\Leftrightarrow \boxed{\frac{\sqrt{\Delta_c}}{24 r_-^3} < \frac{-3qQ}{4r_-}+qQ/r_+} \, ,
\end{eqnarray}
Then the above inequality is equivalent to
\begin{eqnarray}
&&\frac{3qQ}{4r_-}-qQ/r_+<0 ~\left(\frac{r_+}{r_-} < \frac{4}{3}\right)\, , \\
~\cr
&&-\left(18qQr_-^2r_+-24 r_-^3qQ \right) > r_+\sqrt{\Delta_c}\, .
\end{eqnarray}
Then we have
\begin{eqnarray}
&&r_+^2 (2 M - r_+)^2\mu^2-16 M^2 q^2 Q^2  \cr
~\cr
&&+ 20 M q^2 Q^2 r_+ + (l(l+1) - 7 q^2 Q^2) r_+^2>0 \, ,
\end{eqnarray}
which can be simplified as
\begin{equation}\label{poDeltaC}
\mu^2>\frac{q^2Q^2}{r_+^2}\frac{7x^2-20x+16}{(x-2)^2}-\frac{l(l+1)}{r_-^2}.
\end{equation}
One can check that the above condition  is consistent with Eq.~\eqref{deltaMu},
\begin{eqnarray}
&&\frac{q^2Q^2}{r_+^2}\frac{7x^2-20x+16}{(x-2)^2}-\frac{l(l+1)}{r_-^2} \cr
~\cr
&&<\mu^2<\frac{q^2Q^2}{r_+^2}\frac{x^2 (-6 + 7 x)}{8 (-2 + x)^2 (-1 + x)}-\frac{l(l+1)}{r_-^2}.
\end{eqnarray}
Therefore, it is concluded that in case II-II, the superradiantly stable parameter space of the scalar and dyonic black hole system is
\begin{eqnarray}
&&\frac{q^2Q^2}{r_+^2}\frac{7x^2-20x+16}{(x-2)^2}-\frac{l(l+1)}{r_-^2} \cr
~\cr
&&<\mu^2<\frac{q^2Q^2}{r_+^2}\frac{x^2 (-6 + 7 x)}{8 (-2 + x)^2 (-1 + x)}-\frac{l(l+1)}{r_-^2}.
\end{eqnarray}
with
\begin{equation}
\frac{r_+}{r_-}< \frac{4}{3} \left(\frac{Q}{M}>\frac{2\sqrt{6}}{7} \right)\, ,\quad \omega_+<\omega<qQ/r_+.
\end{equation}
Furthermore, we also show the plot of the effective potential in Fig.~\ref{case3}
\begin{figure}[hbtp]
\centering
\includegraphics[width=0.4\textwidth]{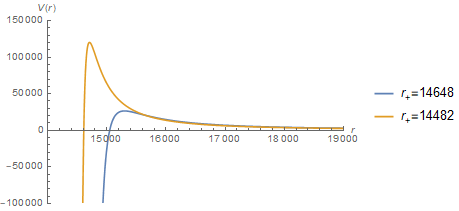}
\caption{Two examples of the effective potential in case II-II. The black hole mass are chosen as
$r_+=14648$ and $ r_+=14482$ for the orange curve and blue curve respectively. The other parameters are chosen as $l=\omega=5, \mu=20, q=15, Q=7000.$  }\label{case3}
\end{figure}
}

\end{itemize}

\section{Summary}
In this paper, we study the superradiant stability of the dyonic black holes in string theory against a charged massive bosonic perturbation. Although the metric of a dyonic black hole is similar to that of a RN black hole and the electric charged black hole in string theory has been proved superradiantly stable, we find that the magnetic field of the dyonic black hole makes the scalar and black hole system more unstable than RN black hole, giving that the system is not superradiantly stale in whole parameter space. Therefore, we discuss three cases for the superradiantly stable region.

For a general dyonic black hole (with the equal electric and magnetic charges) the superradiantly stable parameter region of the system is $\mu^2>\frac{q^2Q^2}{r_+^2}\frac{x^2 (7x-6 )}{8 (x-2)^2 (x-1)}-\frac{l(l+1)}{r_-^2}( x=r_+/M)$. If the ratio between the black hole charge and  the black hole mass satisfies $\frac{Q}{M}<\frac{2\sqrt{6}}{7}$, the superradiantly stable parameter region of the system becomes larger,
 which is $\mu^2 > \frac{q^2Q^2}{r_+^2}\frac{7x^2-20x+16}{(x-2)^2}-\frac{l(l+1)}{r_-^2}$. For the case $\frac{Q}{M}>\frac{2\sqrt{6}}{7}$, the superradiantly stable parameter is $\frac{q^2Q^2}{r_+^2}\frac{x^2 (-6 + 7 x)}{8 (-2 + x)^2 (-1 + x)}-\frac{l(l+1)}{r_-^2}>\mu^2 > \frac{q^2Q^2}{r_+^2}\frac{7x^2-20x+16}{(x-2)^2}-\frac{l(l+1)}{r_-^2}$  and $\omega_+<\omega<qQ/r_+$. For each case, we present a picture to show the superradiantly stable effective potential.

Recently, it has been pointed out that the magnetically charged black holes have some interesting features\cite{Maldacena:2020skw}. They can have long-lived life even their masses are not large. They may have strong magnetic fields, which results in restoring the electroweak symmetry in some regions around them. It is worth noting that  a relevant interesting feature is the Hawking radiation effects are enhanced by the magnetic fields for near extremal black holes. In the near extremal limit, the Hawking emission modes are  in the superradiant region. In this sense, the magnetic fields enhanced the emission of superradiant modes. This is consistent with our result that magnetic fields make black hole system more unstable. It will be interesting to study further about magnetically charged black hole systems.

\acknowledgments
Z.F.M thanks Run-Qiu Yang, H. L\"u and Shou-Long Li for useful discussion and proofreading. J.H.H. is supported by the Natural Science Foundation of Guangdong Province (No.2016A030313444). This work is also partially supported by Guangdong Major Project of Basic and Applied Basic Research (No. 2020B0301030008), Science and Technology Program of Guangzhou (No. 2019050001) and Natural Science Foundation of Guangdong Province (No. 2020A1515010388).


\begin{thebibliography}{99}
\bibitem{raidancecondtion}
Because the magnetic field does not do work on a charged particle, so we assume the superradiant condition for a dyonic black hole is
the same as that for a RN black hole.
\end{thebibliography}
\end{document}